\documentclass[prl,floatfix,twocolumn,showpacs]{revtex4}
\usepackage{amsmath}
\usepackage{amsfonts}
\usepackage{mathrsfs}
\usepackage{epsfig}
\usepackage{graphicx}
\usepackage{dcolumn}

\begin{document}
\title{Complex patterns arise through spontaneous symmetry breaking 
in dense homogeneous networks of neural oscillators
}
\author{Rajeev Singh, Shakti N. Menon and Sitabhra Sinha}
\affiliation{
The Institute of Mathematical Sciences, CIT Campus, Taramani,
Chennai 600113, India.
}
\date{\today}
\begin{abstract}
Recent experiments have highlighted how collective dynamics in
networks of brain regions affect behavior and cognitive function.
In this paper we show that a simple, homogeneous system of
densely connected oscillators representing the aggregate activity of local
brain regions can exhibit a
rich variety of dynamical patterns emerging via spontaneous
breaking of permutation or translational symmetry. 
Our results connect recent experimental findings and suggest that a range of 
complicated activity patterns seen in the brain could be explained 
even without  a full knowledge of its wiring diagram.
\end{abstract}

\maketitle

\newpage
Collective dynamics of coupled oscillators, in
particular, synchronization~\cite{Pikovsky03}, 
is integral to many natural phenomena~\cite{Acebron05} and is
especially important for several biological
processes~\cite{Glass01},
such as brain function~\cite{Engel01,Buzsaki04}. 
While very large-scale synchronization of neuronal activity is considered
pathological, as in epilepsy~\cite{Kandel00}, the brain is capable of
exhibiting a variety of complex spatiotemporal
excitation patterns that may play a crucial role in information
processing~\cite{Singer93}.
Understanding the dynamics of these patterns at the scale of the entire
brain (imaged using techniques such as fMRI) 
is of fundamental importance, as 
interaction between widely dispersed brain regions are responsible for
significant behavioral changes, such as loss of consciousness 
caused by disruption of communication between different areas of the cerebral
cortex~\cite{Lewis12}. 
As detailed
simulation of each individual neuron in the brain 
is computationally expensive~\cite{Markram06}, 
when studying the dynamics of the entire system it is useful to
focus on the network of interactions between brain regions.
It has also been explicitly shown that the collective response of 
a large
number of connected excitatory and inhibitory neurons,
which constitute such 
regions, can be much
simpler than the dynamics of individual neurons~\cite{Vreeswijk96}.
Indeed, each region can be
described using phenomenological models in terms of a few aggregate
variables~\cite{Deco08}.

\begin{figure}
\begin{center}
\includegraphics[width=0.99\linewidth]{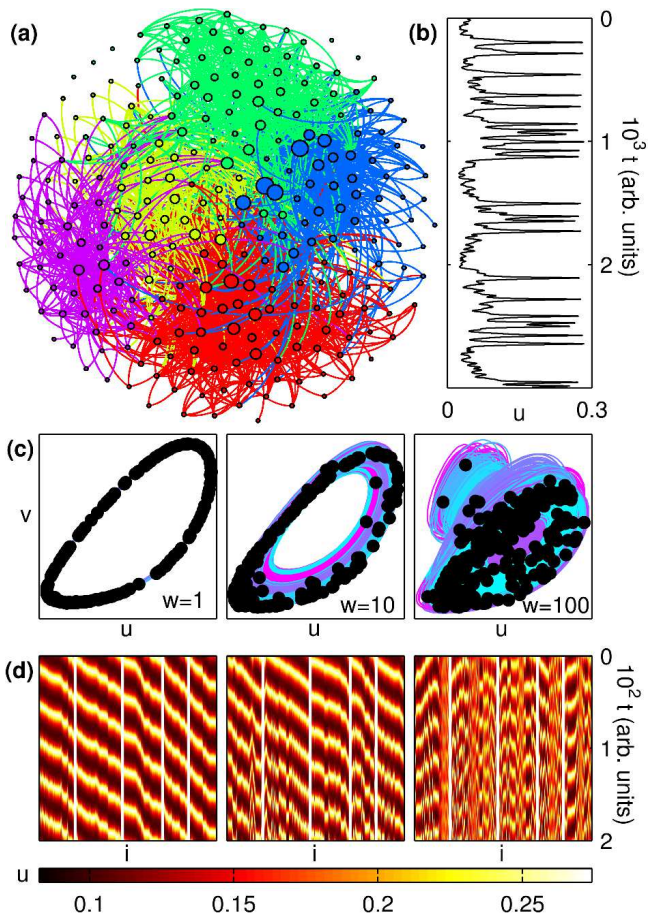}
\end{center}
\caption{(color online). (a) The directed network of connections between
regions of the Macaque brain, adapted from Ref.~\cite{Modha10}.
The size of each node is
proportional to its total degree and the colors distinguish the
modules
(characterized by significantly higher intra-connection density and 
obtained using a partitioning algorithm~\cite{Newman08}).
The
color of each link corresponds to that of the source node. (b) Time
series of the excitatory component of a typical node in this network
with coupling strength $w=500$, where each node is modeled as a
Wilson-Cowan
oscillator.
(c) Phase space projections of the oscillators, obtained for
different coupling strengths, where
the filled circles represent the location of each oscillator at 
a time instant. 
The panels here are scaled individually for better visualization.
(d) Time-series of the excitatory
component $u$, for the corresponding values of $w$
used
in the panels directly above. The nodes $i$ are arranged according to
their modules
(demarcated by white lines).}
\label{fig1}
\end{figure}

%
Using anatomical and physiological data obtained over several decades,
the networks of brain regions for different animals have been
reconstructed~\cite{Scannell95,Modha10}, where the 
individual nodes correspond to large assemblies ($10^3-10^6$) of
neurons~\cite{Palm93}.
The connectivity $C$ (i.e., fraction of realized links) 
of these networks ($C \sim 10^{-1}$) is significantly
higher than that among neurons ($C \sim 10^{-6}$)~\cite{Shepherd03}.
A schematic representation of a network of the Macaque brain regions
(adapted from Ref.~\cite{Modha10}) is shown in Fig.~\ref{fig1}~(a).
The collective activity of such networks can result in complicated
nodal dynamics,
including temporal oscillations at several scales that
are known to be functionally relevant~\cite{Buzsaki04,Schnitzler05}. 
Each of these nodes can be described using neural field models of 
localized neuronal population activity, which can have 
varying mathematical complexity and biological 
realism~\cite{Jansen95}. 
In this paper, we use the well-known and pioneering model proposed
by Wilson and Cowan (WC)~\cite{Wilson72} 
to describe the activity of each brain region.
The resulting temporal patterns of the nodes in the Macaque
network shown in
Fig.~\ref{fig1}~(b)
are reminiscent of experimentally
recorded activity of brain regions~\cite{Buzsaki04}.

The complex collective dynamics observed for the network
at different connection
strengths [Fig.~\ref{fig1}~(c)-(d)] 
can arise from an interplay of several factors, which makes their
analysis difficult. A possible approach to understand the genesis of
these patterns is to focus on the dynamics of the nodes interacting
in the simplified setting of a homogeneous, globally coupled system, which is
an idealization of the densely connected network.
In this paper we show that this simple system 
exhibits an unexpectedly rich variety of complex phenomena, despite
lacking the detailed topological features of brain
networks [e.g., Fig.~\ref{fig1}~(a)], such as
heterogeneity in degree (number of connections per node) and 
modular organization. 
In particular, we show the
existence of novel collective states, including those characterized by
oscillator clusters, where each cluster is distinguished by its
amplitude or frequency.
The occurrence of such clusters is surprising
as each node is identical in
terms of both intrinsic dynamics and connectivity, indicating that the 
homogeneous system of oscillators undergoes
{\em spontaneous symmetry breaking}.
In addition we observe patterns where
the time-series for all oscillators are identical except for a
non-zero phase difference between $n_{cl}$ groups of exactly synchronized
elements which we refer to as ``phase clusters".
On removing a few links
from the network while preserving the structural symmetry of connections we
observe even more dramatic situations such as the appearance of many
($>2$) clusters having different amplitudes. In addition, 
oscillator death, which is seen over a substantial region of
parameter space in the fully connected system, occurs in a
drastically reduced region for such marginally sparse networks.
As the behavior of a large, densely connected system
is effectively identical to that of the 
corresponding mean-field model,
it is remarkable that the dynamical properties of the system considered
here are radically altered in response to extremely minor
deviations from the fully connected situation.

The model we consider comprises a network of $N$ oscillators,
each described by the WC model whose dynamics
results from interactions between an excitatory and an inhibitory
neuronal subpopulation. The average activity of each node $i$ ($u_i,
v_i$) evolves as:
\begin{equation} \label{wc_eqn}
\begin{split}
\tau_u \dot{u}_i &= -u_i + (\kappa_u - r_u u_i)\ {\cal S}_u(u_i^{in}), \\
\tau_v \dot{v}_i &= -v_i + (\kappa_v - r_v v_i)\ {\cal S}_v(v_i^{in}),
\end{split}
\end{equation}
where, $u_i^{in} = c_{uu} u_i - c_{uv} v_i + \sum\nolimits'
(w^{uu}_{ij} u_j - w^{uv}_{ij} v_j) + I^{ext}_{u}$ and
$v_i^{in} = c_{vu} u_i - c_{vv} v_i +
\sum\nolimits'
(w^{vu}_{ij} u_j - w^{vv}_{ij} v_j) + I^{ext}_{v}$
represent the total input to the two
subpopulations respectively.
The time constants and external stimuli for the subpopulations are 
indicated by $\tau_{u,v}$ and $I^{ext}_{u,v}$ respectively, while
$c_{\mu \nu}$ ($\mu, \nu = u,v$)
corresponds to the strength of interactions within and between the
subpopulations of a node.
The interaction strengths are represented by the weight
matrices $W^{\mu \nu} = \{w^{\mu \nu}_{ij}\}$ 
and the summation $\Sigma^{\prime}$ is over all network neighbors.
The function ${\cal S}_{\mu} (z)=[1 + \exp\{-a_{\mu} 
(z - \theta_{\mu})\}]^{-1} + \kappa_{\mu} - 1$
has a sigmoidal dependence on $z$, with
$\kappa_{\mu}=1-[1+\exp(a_{\mu} \theta_{\mu})]^{-1}$.
The parameter values have been chosen such that each isolated node
($W^{\mu \nu} = 0$)
is in the oscillatory regime, viz., 
$a_u=1.3$, $\theta_u=4$,
$a_v=2$, $\theta_v=3.7$,
$c_1=16, c_2=12, c_3=15, c_4=3, r_u=1, r_v=1, \tau_u=8$, $\tau_v=8$,
$I^{ext}_u = 1.25$ and $I^{ext}_v = 0$.
For the homogeneous systems considered here the links will have same strength,
i.e., $w^{\mu \nu}_{ij} = w/k$ 
($\mu, \nu = u,v$ and $i (\neq j) = 1, \ldots, N$),
where $k$ is the degree of a node

The dynamical system (Eq.~\ref{wc_eqn}) is numerically solved using an
adaptive-step Runge-Kutta integration scheme for different system sizes ($N$)
and coupling strengths ($w$).
Linear stability analysis is used to determine the stability of some
of the patterns and identify the associated bifurcations.
The behavior of the system for each set ($w, N$) is analyzed over many ($\sim 100$) randomly chosen initial
conditions. 
We have explicitly verified that our results are robust with respect
to small variations in the parameters.

\begin{figure}
\begin{center}
\includegraphics[width=0.99\linewidth]{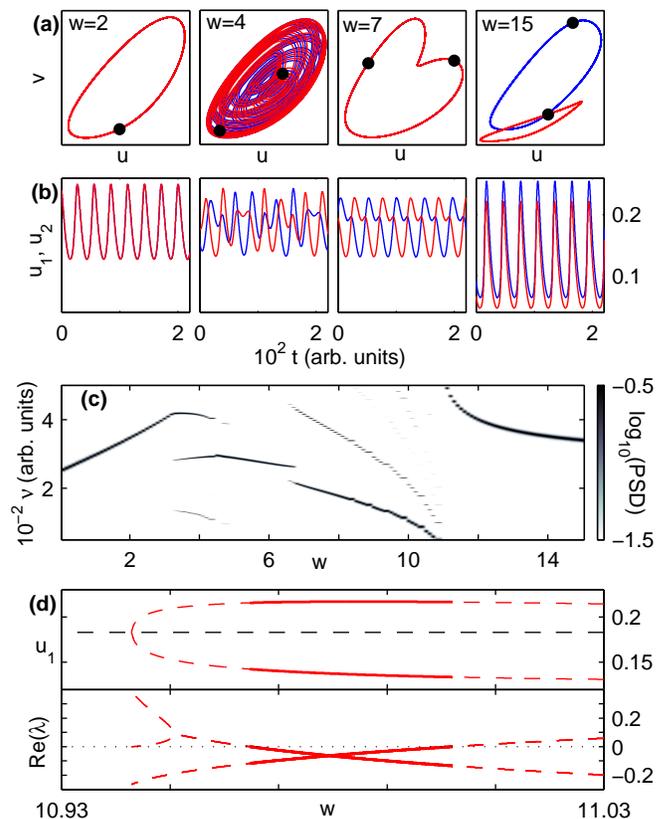}
\end{center}
\caption{(color online). Collective dynamics of a system of two
coupled WC oscillators. (a) Phase space projections of the
trajectories and (b) time-series for each oscillator
showing exact synchronization (ES, for coupling $w=2$),
quasiperiodicity (QP, $w = 4$), anti-phase synchronization (APS,
$w=7$) and inhomogeneous in-phase synchronization (IIS, $w=15$).
The filled circles represent the location of each oscillator in phase
space at a time instant.  The panels in (a) are scaled individually
for better visualization.
(c) Power-spectral density (PSD) of the time-series for the $u$
component of each oscillator, revealing the dominant frequencies
as a function of $w$.
(d) All fixed points of the system (upper panel) and the real parts of
the eigenvalues corresponding to the heterogeneous fixed points (lower
panel) showing the transitions between APS and IIS regimes. Solid
(broken) lines represent stable (unstable) solutions.
}
\label{fig2}
\end{figure}

We first examine the collective dynamics of a pair of coupled
oscillators
($N=2$) as a function of the interaction strength between them.
Fig.~\ref{fig2}~(a)-(b) show that while exact synchronization (ES) of
oscillator dynamics occurs at weak coupling ($w \lesssim 3.2$), a
state of
anti-phase synchronization (APS) is observed at higher values of $w$
($4.4 \lesssim w \lesssim 11$). For intermediate $w$, the co-existence
of
the dominant frequencies corresponding to ES and APS states
[Fig.~\ref{fig2}~(c)]
indicates that the quasi-periodic behavior observed in this regime can
be
interpreted as arising through competition between the mechanisms
responsible for the above two states. 
At $w \sim 11$, the system undergoes
spontaneous
symmetry-breaking , eventually giving rise to inhomogeneous in-phase
synchronization (IIS), characterized by different phase-space
projections
and distinct amplitudes for the time-series of each oscillator
[Fig.~\ref{fig2}~(a)-(b), last panel]. The nature of the transition
from
APS to IIS is made explicit in Fig.~\ref{fig2}~(d) [top panel], where the fixed
points of
one of the oscillators, obtained using numerical root finding, are shown
over a range of $w$.
At $w \approx 10.943$, a pair of heterogeneous
unstable
solutions related by permutation symmetry, corresponding to an
inhomogeneous steady-state (ISS), emerge from a homogeneous
unstable solution, beyond which all three solutions coexist.
Thus,
spontaneous symmetry breaking appears to arise in the system through a
subcritical pitchfork bifurcation, with the number of positive
eigenvalues
corresponding to the homogeneous solution decreasing by unity (not
shown).
The ISS is stable over a very small
range, $10.964 \lesssim w \lesssim 11.002$, as seen from their
corresponding eigenvalues in Fig.~\ref{fig2}~(d) [lower panel]. Note
that
stability is lost on either end of this interval through
supercritical Hopf bifurcations. For $w \gtrsim 700$, both
oscillators converge to the inactive state $u_i=v_i=0, \forall i$,
corresponding to amplitude death (AD, not shown).

\begin{figure}
\begin{center}
\includegraphics[width=0.99\linewidth]{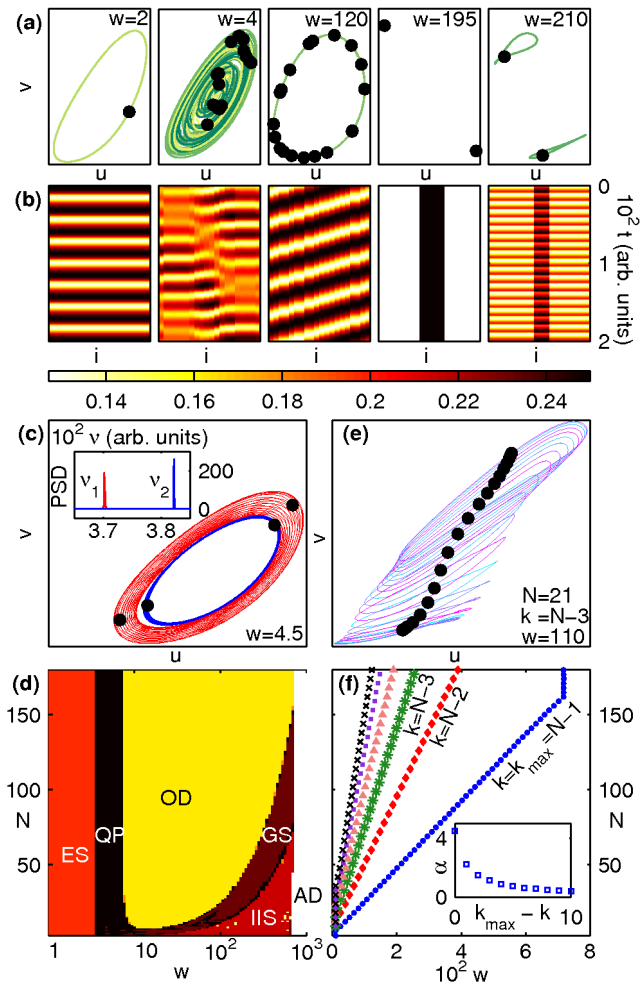}
\end{center}
\caption{(color online).
Collective dynamics of $N$ densely connected WC oscillators. (a) Phase
space
projections of the trajectories and (b) time-series for $N=20$
globally
coupled oscillators, showing exact synchronization (ES, $w=2$),
quasiperiodicity (QP, $w = 4$), gradient synchronization (GS,
$w=120$),
inhomogeneous steady-state (ISS, $w=195$) and inhomogeneous in-phase
synchronization (IIS, $w=210$). 
The panels in (a) are scaled individually for
better visualization. (c) Phase space projections of the different
oscillators for $w = 4.5$, which form two clusters with frequencies
$\nu_1$
and $\nu_2$, indicated by the power-spectral density (PSD, inset). (d)
Phase diagram for $N$ WC oscillators globally coupled with strength
$w$,
indicating areas where the majority ($> 50\%$) of initial conditions
result in
ES, QP, GS, IIS, oscillator death (OD) and
amplitude death (AD). Note that the $w$-axis is logarithmic. 
(e) As the degree
$k$, i.e., the number of links per node, deviates slightly from the
globally
coupled case ($k_{max}$ $= N-1$) to $N-3$, the trajectories of the IIS
state
split into many ($\sim N$) distinct projections ($N=21, w = 110$). (f)
The
OD region in (d) shrinks rapidly with the number of removed links, as
seen
from the slope of the upper boundary of OD (inset).
}
\label{fig3}
\end{figure}
Increasing the number of oscillators, we observe that while the
patterns
seen for a pair of coupled oscillators, namely ES, QP, ISS, IIS, APS and AD
persist (first four shown in Fig.~\ref{fig3}~(a)-(b) for
$N=20$)~\cite{note1},
qualitatively different states also emerge.
As mentioned earlier, a
new class of patterns characterized by the existence of phase
clusters appears. The most robust of these, referred to as gradient
synchronization (GS), has $n_{cl} \sim N$. 
Another new pattern
comprises
two oscillator clusters, each characterized by a unique frequency
[Fig.~\ref{fig3}~(c)]. This constitutes a dramatic instance of
spontaneous
breaking of permutation symmetry, as the oscillators are intrinsically
indistinguishable for this
completely homogeneous system. Thus, the appearance of
multiple
frequencies in a dynamical network need not imply
heterogeneity
in connectivity or node properties.

A third new pattern is a homogeneous steady state referred to as
oscillator
death (OD), in which the individual nodes have the same
time-invariant,
non-zero activity. This dynamical state appears over a large region in
$(w,N)$-space as seen in the phase diagram, Fig.~\ref{fig3}~(d). To
identify and segregate the regimes in this diagram, we use several
order
parameters. The mean of the oscillation amplitude, measured as the
variance
($\sigma^2$) with respect to time of one of the WC variables, $v$,
averaged over all the nodes $\langle \sigma^2_t (v_i) \rangle_i$, is
zero
for all the non-oscillating states AD, OD and ISS. These are further
distinguished by using the mean and variance with respect to all nodes
of
the time-averaged $v$, i.e., $\langle \langle v_i \rangle_t \rangle_i$
(=0 for AD) and $\sigma^2_i (\langle v_i \rangle_t)$ (=0 for OD and
AD). To
distinguish between the oscillating patterns, we consider the mean
coherence, measured as $\langle \sigma^2_i (v_i) \rangle_t$, and the
total
space occupied by all the trajectory projections $\Delta$, as measured
by
the number of non-zero bins of their histogram in ($u,v$)-space. ES is
characterized by $\langle \sigma^2_i (v_i) \rangle_t = 0$ and IIS by
$\sigma^2_i (\langle v_i \rangle_t) > 0$. The remaining patterns, GS
and QP,
are distinguished by $\Delta \sim 0$ for GS. Note that $\langle
\rangle_t$
and $\langle \rangle_i$ represent averaging over time and all nodes,
respectively. In practice, different regimes are characterized by
thresholds whose specific values do not affect the qualitative nature
of
the results.

As a first step towards extending the results seen for the globally
coupled
system to brain networks of the type shown in Fig.~\ref{fig1}~(a), we
have
investigated the consequences of gradually decreasing the connection
density. To ensure that the degree reduction preserves as many of the
existing symmetries as possible, we arrange the nodes on a circle and
sequentially remove connections between nodes placed furthest apart.
In addition to preserving degree homogeneity, this ensures that every node has
the
same neighborhood structure. 
As we deviate from the global coupling limit, we observe patterns similar
to those shown in Fig.~\ref{fig3}~(a-d), although the precise form of
the attractors may differ and it is now the translational symmetry
that is being spontaneously broken. 
For example, as seen in Fig.~\ref{fig3}~(e), a
reduction of just 2 links per node causes the trajectory in the IIS
state
to split into many more ($\sim N$) projections than seen for the fully
connected case ($\sim 2$). Also, while the phase diagram of the system
remains qualitatively unchanged when the degree is decreased from
$k_{max}
= N-1$, there is a dramatic quantitative reduction in the area
corresponding to OD [Fig.~\ref{fig3}~(f)] even with the reduction of
one link per node. This is surprising, as one would expect that
a marginal
deviation from the global coupling limit in large systems will not
result
in a perceptible change from the mean-field behavior.

Our result that weakening connections between nodes of a network can
increase coherence in collective activity (viz., observation of ES at
low
$w$) suggests an intriguing relation between two recent experimental
findings: (i) anaesthetic-induced loss of consciousness occurs
through
the progressive disruption of communication between brain
areas~\cite{Lewis12} and (ii) functional connectivity networks
reconstructed
from EEG data become increasingly dense with the development of fatigue in
sleep-
deprived subjects~\cite{Kar11}. The latter study finds that the 
onset of sleep is accompanied by an increase in
the degree of synchronization between brain areas,
while the former result implies that the interaction strengths between
these
areas will concurrently get weaker. Although it may appear
counter-intuitive that decreased coupling strength would result in increased
synchronization, our findings illustrate that these results
are
not incompatible.

Another important implication of this study follows from our
demonstration
that systems with simple connection topology are capable of exhibiting
very
rich dynamical behavior. In particular, many of the patterns seen in
our
simulations of the network of Macaque brain regions (Fig.~\ref{fig1})
resemble those observed using much simpler connectivity schemes
(Fig.~\ref{fig3}). Hence, patterns seen in complex systems that are
often
attributed to their non-trivial connection structure, may in fact be
independent of the details of the network architecture. In light of
recent
studies that investigate the collective dynamics of networks reconstructed
from
real-life data (e.g., Ref.~\cite{Haimovici13}), our results imply that
caution should be exercised in linking observed features of a system
to
specific properties of the underlying network as, in some cases,
simpler
topologies may reproduce similar patterns.
Thus, our findings provide a baseline for future studies on the
specific role of the detailed aspects (degree heterogeneity,
modular architecture, etc.) of brain networks on their collective
dynamics.

To conclude, we have shown that
the collective dynamics of a homogeneous system of oscillators, motivated
by mesoscopic descriptions of brain activity, exhibits 
spontaneous symmetry breaking that gives rise to several novel
patterns.
Despite
preserving the structural symmetry of connections, a marginal increase
in
the network sparsity, corresponding to an extremely small deviation
from the
mean-field, unexpectedly changes the robustness of certain patterns.
Our results suggest that some of the complicated activity patterns
seen in the brain can be explained even without  
complete knowledge of its wiring diagram.

We thank Raghavendra Singh, Purusattam Ray and Gautam Menon 
for helpful discussions.
We thank IMSc for providing access to the ``Annapurna" supercomputer.
This research was supported in part by the IMSc Complex Systems Project.

\pagebreak
\begin{table*}
{\large \bf SUPPLEMENTARY MATERIAL}
\end{table*}
\setcounter{figure}{0}
\renewcommand\thefigure{S\arabic{figure}}  
\renewcommand\thetable{S\arabic{table}}  
\begin{figure*}
\begin{center}
\includegraphics[width=0.9\linewidth]{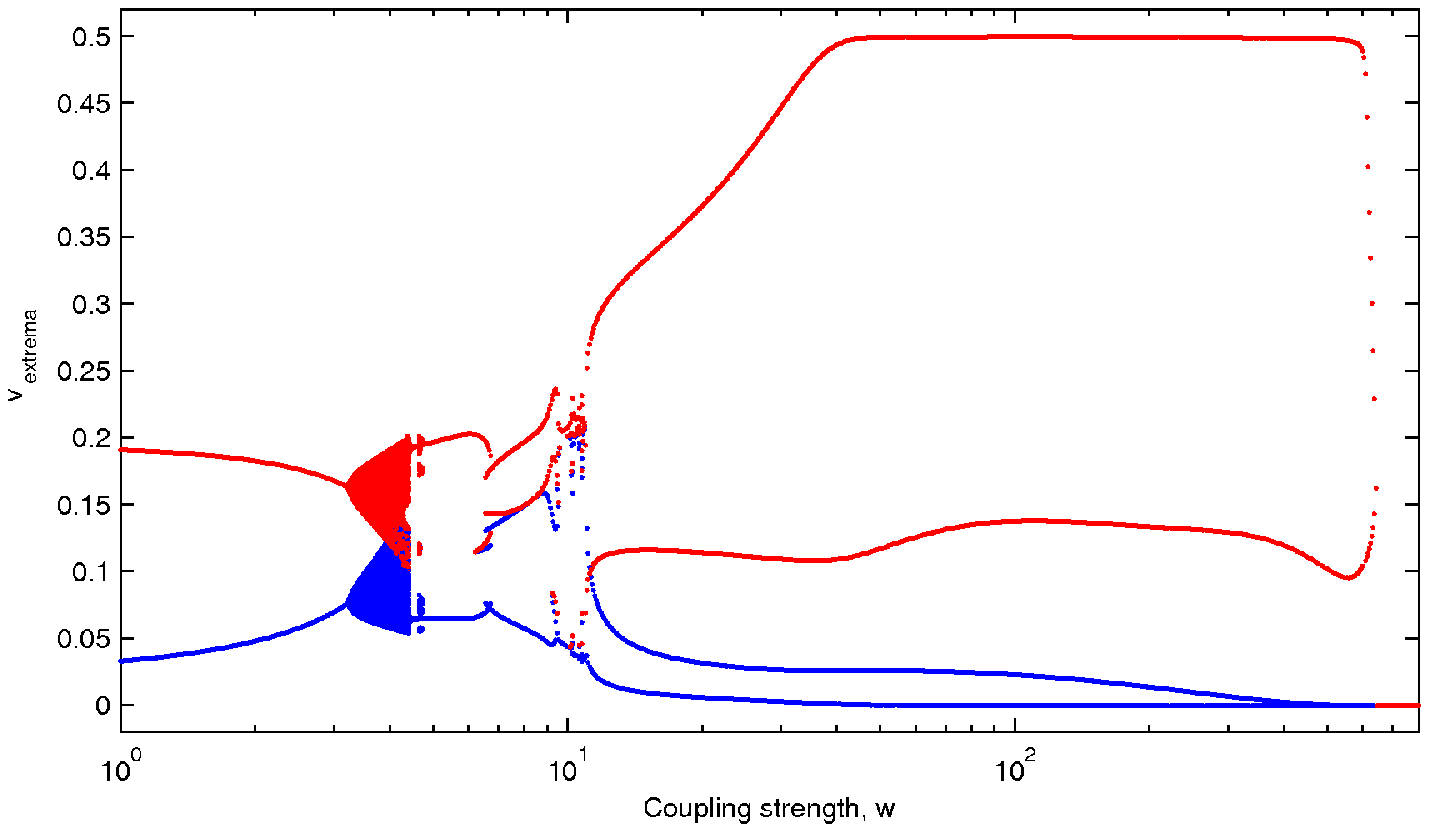}
\end{center}
\caption{Bifurcation diagram for a pair of coupled WC oscillators with
coupling strength $w$ as the bifurcation parameter,
obtained for a set
of 20 random initial conditions (i.c.). Red dots represent the maxima 
of the inhibitory components $v$ for each i.c., 
while blue dots represent the corresponding minima.
}
\label{figs1}
\end{figure*}

\begin{figure*}
\begin{center}
\includegraphics[width=0.9\linewidth]{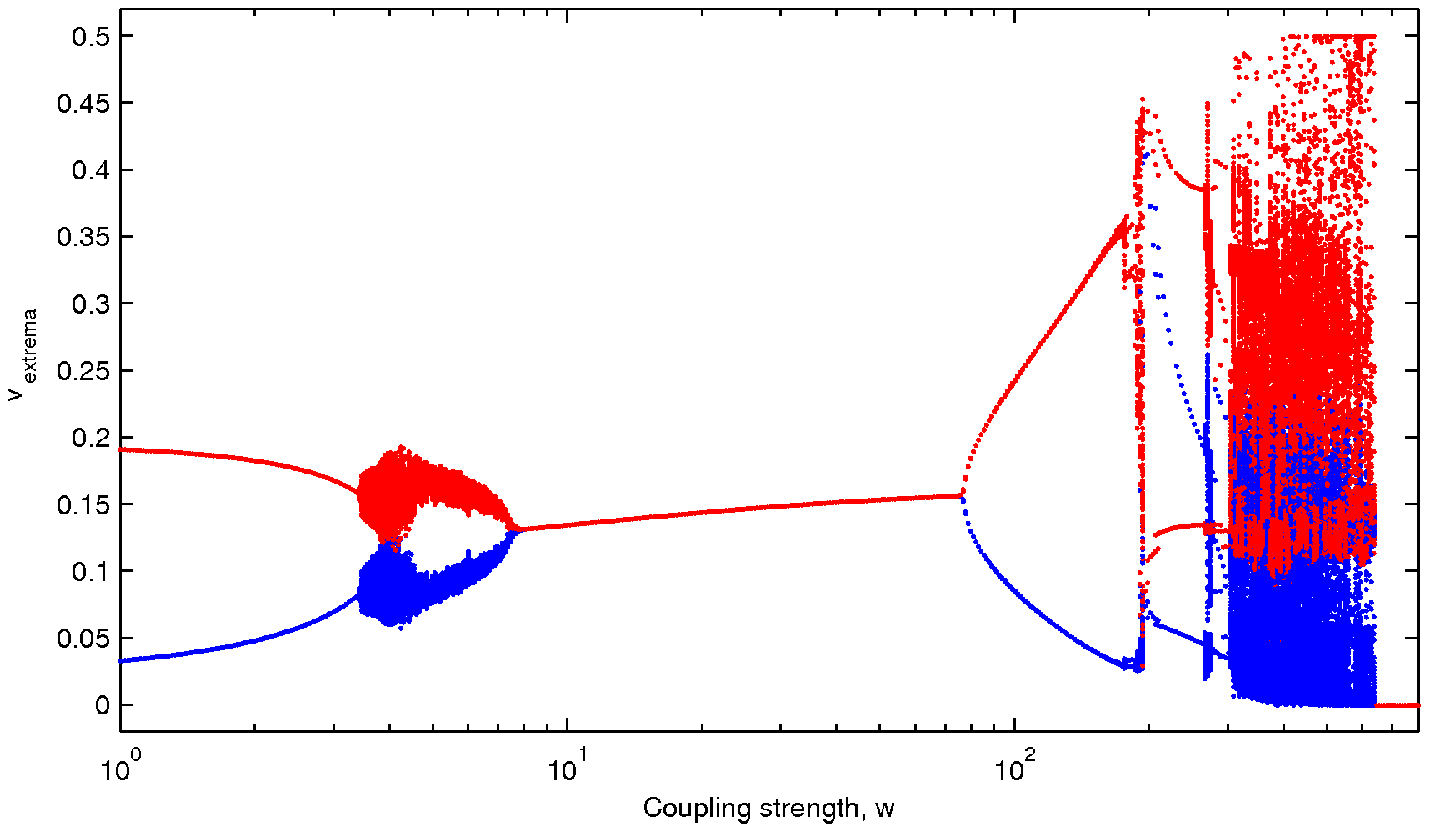}
\end{center}
\caption{Bifurcation diagram for $N=20$ globally coupled WC
oscillators with coupling strength $w$ as the bifurcation parameter,
obtained for a set
of 20 random initial conditions (i.c.). Red dots represent the maxima 
of the inhibitory components $v$ for each i.c.,
while blue dots represent the corresponding minima.
}
\label{figs2}
\end{figure*}

\begin{figure*}
\begin{center}
\includegraphics[width=0.99\linewidth]{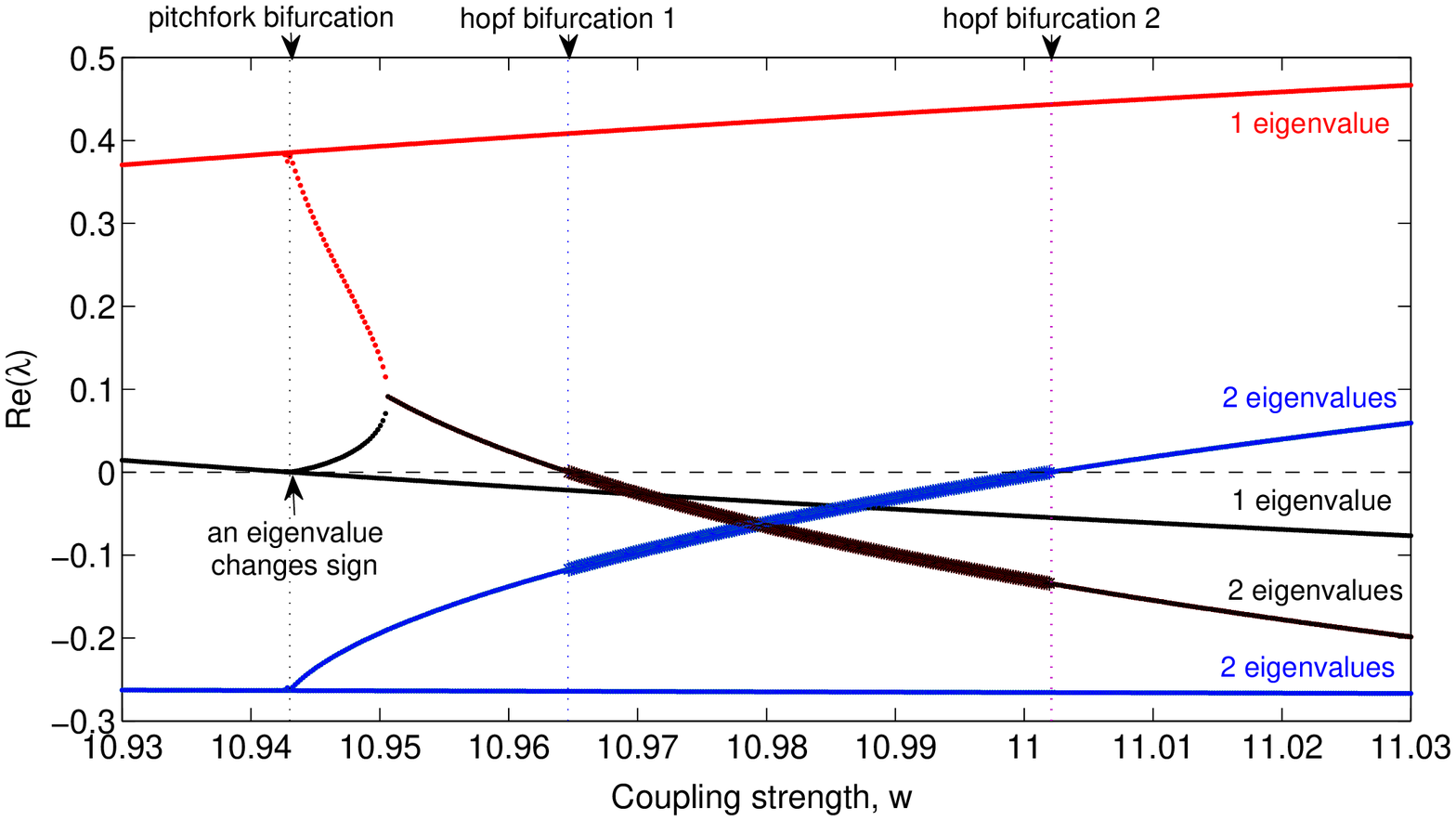}
\end{center}
\caption{Real parts of
the eigenvalues of all the fixed points for a pair of coupled WC
oscillators, shown as a function of coupling strength $w$ in the
neighborhood of the transition between APS and IIS regimes. 
The vertical dotted lines indicate the locations where different bifurcations
occur in this range of $w$.
Thick lines between the two Hopf bifurcations represent stable
solutions. Three of the branches shown correspond to a pair of
eigenvalues, as indicated in the figure.
}
\label{figs3}
\end{figure*}

\begin{table*}
\caption{\label{tab:OrderParameter}Order parameters used for
identifying the different dynamical regimes of a homogeneous network
of WC oscillators (as explained in the main text).}
\begin{ruledtabular}
\begin{tabular}{cccccc}
&&&&& \\
Pattern &
$\langle \sigma^2_t (v_i) \rangle_i = 0$ &
$\langle \langle v_i \rangle_t \rangle_i = 0$ &
$\sigma^2_i (\langle v_i \rangle_t) = 0$ &
$\langle \sigma^2_i (v_i) \rangle_t = 0 $ &
$\Delta > 0$ \\
&&&&& \\
\hline
&&&&& \\
AD  & \checkmark & \checkmark & \checkmark & \checkmark & \\
OD  & \checkmark &  & \checkmark & \checkmark & \\
ISS & \checkmark &  &  &  & \\
ES  &  &  & \checkmark & \checkmark & \\
QP  &  &  &  &  & \checkmark \\
IIS &  &  &  &  & \\
GS  &  &  & \checkmark &  & \\
&&&&& \\
\end{tabular}
\end{ruledtabular}
\end{table*}


\begin{thebibliography}{}
\bibitem{Pikovsky03}
{A. Pikovsky, M. Rosenblum, and J. Kurths,
{\em Synchronization} (Cambridge University Press,
Cambridge, England, 2003).}

\bibitem{Acebron05}
{J.~A. Acebr\'{o}n {\em et al.}, Rev. Mod. Phys. {\bf 77}, 137
(2005).}

\bibitem{Glass01} 
{L. Glass, Nature (London) {\bf 410}, 277 (2001);}
{M.~U. Gillette and T.~J. Sejnowski, Science {\bf 309}, 1196
(2005);}
{R. Singh {\em et al.}, Phys. Rev. Lett. {\bf 108}, 068102 (2012).}

\bibitem{Engel01}
{A.~K. Engel, P. Fries and W. Singer,
Nature Rev. Neurosci. {\bf 2}, 704 (2001);}
{M.~I. Rabinovich, P. Varona, A.~I. Selverston and H.~D.~I. Abarbanel,
Rev. Mod. Phys. {\bf 78}, 1213 (2006).}

\bibitem{Buzsaki04}
{G. Buzs\'{a}ki and A. Draguhn, Science {\bf 304}, 1926 (2004).}

\bibitem{Kandel00}
{E.~R. Kandel, J.~H. Schwartz and T.~M. Jessell,
{\em Principles of Neural Science} (McGraw-Hill,
New York, 4th edition, 2000).}

\bibitem{Singer93}
{W. Singer, Ann. Rev. Physiol. {\bf 55}, 349 (1993).}

\bibitem{Lewis12} 
L.~D. Lewis {\em et al.}, Proc. Natl. Acad. Sci. USA {\bf 109}, E3377
(2012).

\bibitem{Markram06}
H. Markram, Nature Rev. Neurosci. {\bf 7}, 153 (2006);
{C. Zhou, L. Zemanov\'{a}, G. Zamora, C.~C. Hilgetag and J. Kurths,
Phys. Rev. Lett. {\bf 97}, 238103 (2006).}

\bibitem{Vreeswijk96} 
{C. van Vreeswijk and H. Sompolinsky, Science {\bf 274}, 1724 (1996).}

\bibitem{Deco08}
{G. Deco, V.~K. Jirsa, P.~A. Robinson, M. Breakspear and K. Friston,
PLoS Comput. Biol. {\bf 4}, e1000092 (2008).}

\bibitem{Scannell95}
{J.~W. Scannell, C. Blakemore and M.~P. Young,
J. Neurosci. {\bf 15}, 1463 (1995);}
{P. Hagmann, L. Cammoun, X. Gigandet, R. Meuli, C. J.
Honey, V. J. Wedeen, and O. Sporns, PLoS Biol. {\bf 6}, e159
(2008).}

\bibitem{Modha10} {D.~S. Modha and R. Singh, Proc. Natl. Acad. Sci.
USA {\bf 107}, 13485 (2010).}

\bibitem{Palm93} 
G. Palm, Hippocampus {\bf 3}, 219 (1993);
C. Johansson and A. Lansner, Neural Networks {\bf 20}, 48 (2007).

\bibitem{Shepherd03}
{G.~M. Shepherd (Ed.), {\em The Synaptic Organization of the Brain}
(Oxford University Press, New York, 4th edition, 2003);
R. Morris, in {\em The Hippocampus Book} (Eds. P. Andersen {\em et
al.}) (Oxford University Press, Oxford, 2007).}

\bibitem{Newman08}
{E.~A. Leicht and M.~E.~J. Newman, Phys. Rev. Lett. {\bf 100}, 118703
(2008).}

%

\bibitem{Schnitzler05}
{A. Schnitzler and J. Gross, Nature Rev. Neurosci. {\bf 6}, 285
(2005);}
{P.~J. Uhlhaas and W. Singer, Nature Rev. Neurosci. {\bf 11}, 100
(2010).}

\bibitem{Jansen95}
B.~H. Jansen and V.~G. Rit, Biol. Cybern. {\bf 73}, 357 (1995);
C.~J. Honey, R. Kotter, M. Breakspear and O. Sporns, Proc. Natl. Acad.
Sci. USA {\bf 104}, 10240 (2007);
{F. Marten, S. Rodrigues, O. Benjamin, M.~P. Richardson and J.~R.
Terry, Phil. Trans. R. Soc. A {\bf 367}, 1145 (2009).}

\bibitem{Wilson72} {H.~R. Wilson and J.~D. Cowan,
Biophys. J. {\bf 12}, 1 (1972);}
{A. Destexhe and T.~J. Sejnowski, Biol. Cybern. {\bf 101}, 1 (2009).}

\bibitem{note1}
{Note that APS, which for $N>2$ has a very small basin of attraction, 
is a ``phase cluster" state for which $n_{cl} = 2$.}

\bibitem{Kar11}
S. Kar, A. Routray and B.~P. Nayak,
Clin. Neurophysiol. {\bf 122} 966 (2011);
C.~J. Chu {\em et al.}, J. Neurosci. {\bf 32} 2703 (2012).

\bibitem{Haimovici13}
{A. Haimovici, E. Tagliazucchi, P. Balenzuela and D.~R. Chialvo, Phys.
Rev. Lett. {\bf 110}, 178101 (2013).}

\end{thebibliography}
\end{document}